\documentclass[twocolumn,english,pre,showpacs]{revtex4-1}
\usepackage[colorlinks=true,urlcolor=blue,citecolor=blue,linkcolor=blue]{hyperref} 
\usepackage[T1]{fontenc}
\usepackage[latin9]{inputenc}
\usepackage{amssymb}
\usepackage{graphicx}
\usepackage{amsmath,color}
\usepackage{mathrsfs}
\usepackage{float}
\usepackage{indentfirst}
\usepackage{braket}

\makeatletter

\def\journal #1, #2, #3, 1#4#5#6{{\sl #1~}{\bf #2}, #3 (1#4#5#6) }

\def\eqa{\begin{eqnarray}}
\def\eea{\end{eqnarray}}
\newcommand{\eq}{\begin{equation}}
\newcommand{\ee}{\end{equation}}

\newcommand{\Eq}[1]{Eq.~(\ref{#1})}

\makeatother

\usepackage{babel}

\begin{document}

\title{Stochastic series expansion simulation of the $t$-$V$ model}

\author{Lei Wang, Ye-Hua Liu and Matthias Troyer}
\affiliation{Theoretische Physik, ETH Zurich, 8093 Zurich, Switzerland}

%\affiliation{$^{1}$Beijing National Lab for Condensed Matter Physics and Institute
%of Physics, Chinese Academy of Sciences, Beijing 100190, China }

\begin{abstract}
We present an algorithm for the efficient simulation of the half-filled spinless $t$-$V$ model on bipartite lattices, which  combines the stochastic series expansion method with  determinantal quantum Monte Carlo techniques widely used in fermionic simulations. The algorithm scales linearly in the inverse temperature, cubically with the system size and is free from the time-discretization error. We use it to map out the finite temperature phase diagram of the spinless $t$-$V$ model on the honeycomb lattice and observe a suppression of the critical temperature of the charge density wave phase in the vicinity of a fermionic quantum critical point. 
\end{abstract}


%71.10.Fd Lattice fermion models,
%71.27.+a Strongly correlated electron systems, 
%02.70.Ss quantum Monte Carlo

\maketitle

\section{Introduction}
%SSE 
%Developing even more powerful numerical algorithms lies in the heart of computational study of strongly correlated systems. In region of quantum Monte Carlo methods, 
The stochastic series expansion (SSE) method~\cite{Sandvik:1991tn} is an efficient and versatile numerical method for unbiased simulations of quantum many-body systems. It performs stochastic sampling of the Taylor series expansion of the partition function. Combined with nonlocal Monte Carlo updates~\cite{1999PhRvB..5914157S, Syljuasen:2002hw, Alet:2005cv}, SSE is the method of choice~\footnote{For systems with large diagonal terms (such as spins in magnetic field or soft core bosons with onsite energy), the worm algorithm in the path-integral representation is advantageous~\cite{Prokofev:1998tc}} for the simulation of unfrustrated quantum-spin models and hard-core bosons~\cite{Sandvik:kca, Kaul:2013ika}. 

%While the CT-QMC methods starts from of the partition function in the interaction picture. 
%Typically, the SSE method is applies to quantum spin systems where different terms are of the same magnitude, while the path-integral formalism is better  with big on site energy such as the bose-Hubbard models. 
%det-QMC
The fermion sign problem prevents a direct application of the SSE method to fermionic systems beyond one spatial dimension~\cite{Sandvik:1992wb}. Fermion exchange processes in higher dimension give rise to a fluctuating sign and thus prevent efficient Monte Carlo sampling. To alleviate the fermion sign problem, a common choice for fermionic simulations is the determinantal quantum Monte Carlo (QMC) approach, in both the traditional discrete time formulation~\cite{Blankenbecler:1981vj} and more recent continuous-time formulations~\cite{1999PhRvL..82.4155R, Rubtsov:2005iw, Iazzi:2015hi, 2015PhRvB..91w5151W}. Both approaches map an interacting fermion problem to noninteracting fermions subjected to imaginary-time-dependent actions. By tracing out these free fermions, one is able to resum a factorially large number of fermion exchange processes into a single determinant. Although in general this resummation does not completely solve the fermion sign problem, in special cases the determinant has a definite sign due to symmetry. For example, the determinant can be nonnegative either due to the time-reversal symmetry~\cite{PhysRevC.48.1518, Hands:2000kq, Wu:2005im} or due to a more recently discovered split orthogonal group symmetry~\cite{Wang:2015hm}. %Furthermore, by going to the Majorana representation~\cite{Li:2015jf}, one can identify even broader class of sign-free problems~\cite{MajoranaPositivity, MajoranaTRS}. 

%explain the main idea
In this paper, we present an algorithm combining the SSE and determinantal QMC techniques. When applicable, the method can avoid the fermion sign problem in the conventional SSE approach~\cite{Sandvik:1991tn} and outperform conventional determinantal approaches~\cite{Blankenbecler:1981vj, 1999PhRvL..82.4155R, Rubtsov:2005iw, Iazzi:2015hi, 2015PhRvB..91w5151W}. After presenting the algorithm, we will use it to map out the finite temperature phase diagram of the spinless $t$-$V$ model on the honeycomb lattice, whose  Hamiltonian reads  
\begin{equation}
  \hat{H} = \sum_{\braket{ i, j }} - t \left(\hat{c}_i^{\dagger}\hat{c}_j + \hat{c}_j^{\dagger} \hat{c}_i\right) + V\left(\hat{n}_{i}-\frac{1}{2}\right) \left(\hat{n}_{j}-\frac{1}{2}\right),  
 \label{eq:model}
\end{equation}
where $\hat{c}_{i}$ is the fermion annihilation operator on site $i$, $\hat{n}_{i}=\hat{c}_{i}^{\dagger}\hat{c}_{i}$ is the occupation number operator. $V>0$ denotes the repulsive interaction. On bipartite lattices the sign of hopping amplitude $t$ is irrelevant. For definiteness we let $t>0$ in the following discussion. 

\begin{figure}
 \includegraphics[width=\columnwidth]{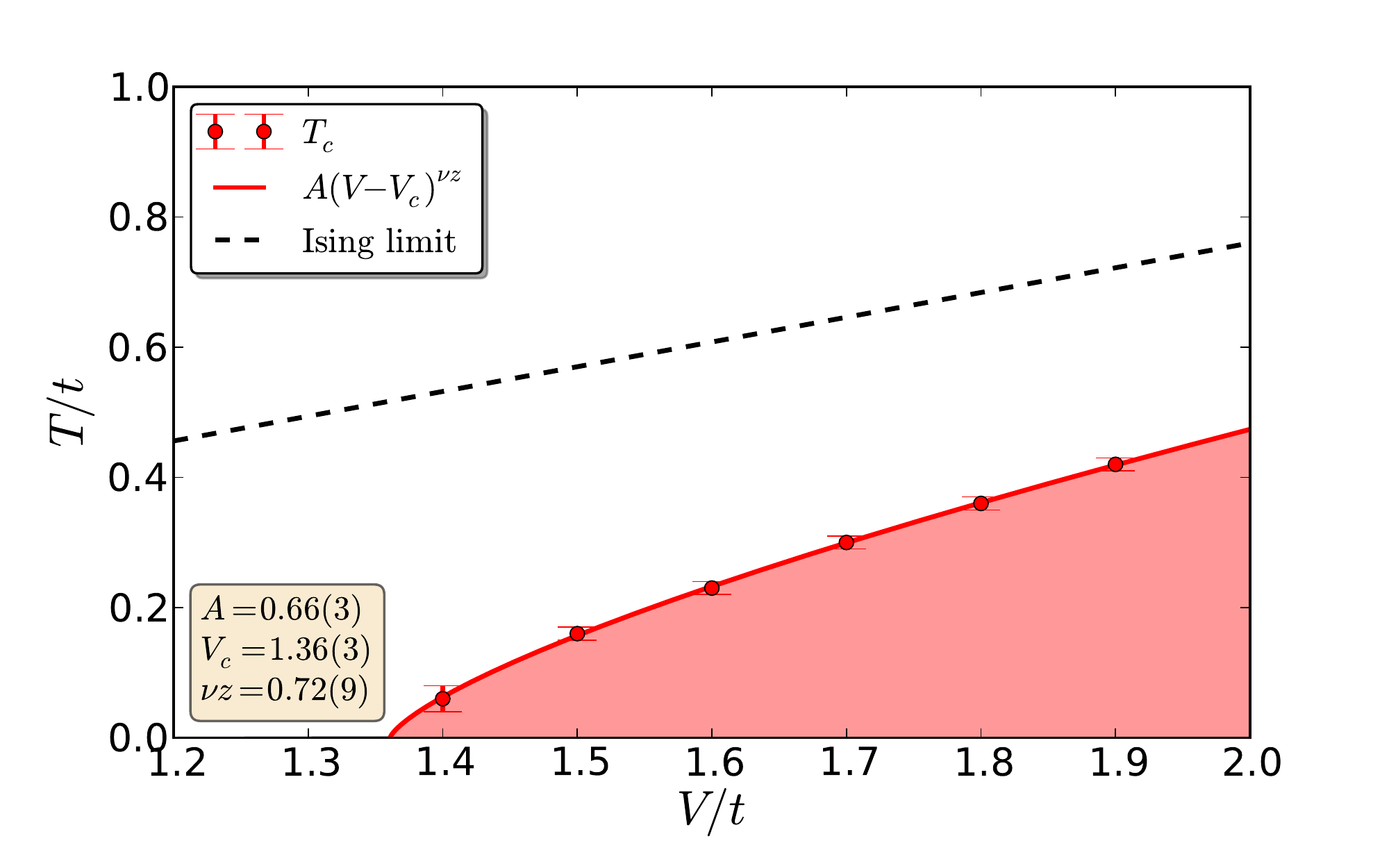}
 \caption{Phase diagram of the model (\ref{eq:model}) on the honeycomb lattice. Shaded region is in the CDW phase. The critical temperature $T_{c}$ approaches to zero at the quantum critical point between the CDW and Dirac-semimetal states. The red solid line is a fit of the critical temperature to $T_{c}= A(V-V_{c})^{\nu z}$. The dashed black line indicates the critical temperature of the Ising limit $T_{c}=0.38V$~\cite{1967RPPh...30..615F} applies to the strong-coupling limit $V\gg t$.}
\label{fig:Phasediag}
\end{figure}

%facts about the model. 
In one dimension the model (\ref{eq:model}) can be mapped to a spin-$1/2$ XXZ model through a Jordan-Wigner transformation, which allows for efficient SSE simulations. However,  SSE simulations in higher dimensions suffer from a fermion sign problem.  The model also suffers from a severe fermion sign problem even in the conventional determinantal QMC method~\cite{Scalapino:1984wz, Gubernatis:1985wo}. The meron-cluster method solves the sign problem for $V\ge2t$~\cite{Anonymous:xaWVK-gC}. For general $V>0$  the sign problem has recently been solved by the continuous-time interaction expansion method~\cite{Rubtsov:2005iw} using the Fermi bag idea~\cite{Huffman:2014fj} \footnote{Later it was realized that the model (\ref{eq:model}) is \emph{naturally} sign-problem free in the continuous-time interaction expansion methods \cite{Wang:2014iba, 2015PhRvB..91w5151W}.} and in the discrete-time method~\cite{Blankenbecler:1981vj} by using the Majorana fermion representation~\cite{Li:2015jf}. These two solutions have been unified by revealing the underlying Lie group structure of the determinantal QMC methods~\cite{Wang:2015hm}, therefore provides a useful guiding principle for sign-free QMC simulations. Very recently,  Ref.~\cite{MajoranaPositivity} further extends these solutions via considering  Majorana reflection positivity conditions.  Reference~\cite{MajoranaTRS} systematizes the idea of~\cite{Li:2015jf} via classifying a set of anti-unitary and mutually anti-commuting operations in the Majorana basis. Based on these developments~\cite{Anonymous:xaWVK-gC, Huffman:2014fj, Li:2015jf}, the Ising phase transition of staggered fermions~\cite{Anonymous:UJp124el,Anonymous:WZs4V-GU}, the fermionic quantum critical point of the model on honeycomb and $\pi$-flux lattices~\cite{Wang:2014iba, 2015PhRvB..91w5151W, Li:2015cwb} and the R\'enyi entanglement entropy across thermal and quantum phase transitions~\cite{Wang:2014ir, Broecker:2015vk} have been studied.

The SSE method presented in this paper is a further algorithmic development. It solves the sign problem of the model (\ref{eq:model}) for the whole range of repulsive interaction $V>0$,  is free from the time-discretization errors and allows for more efficient implementations compared to  previous simulations~\cite{Wang:2014iba, Li:2015cwb, 2015PhRvB..91w5151W}. Using this algorithm we obtain unbiased results for over a thousand lattice sites and map out the finite-temperature phase diagram on the honeycomb lattice as shown in Fig.~\ref{fig:Phasediag}. The system is in a staggered charge-density-wave (CDW) phase in the shaded strong-coupling and low-temperature region. The CDW critical temperature approaches zero at a fermionic quantum critical point found in Refs.~\cite{Wang:2014iba, 2015PhRvB..91w5151W, Li:2015cwb}.    

\section{The determinental SSE algorithm}
%\subsection{Basic algorithm}
Observing the Hamiltonian (\ref{eq:model}) to be a summation of local terms defined on lattice bonds, we introduce the bond index $b=\braket{i,j}$ and hopping operator for each bond $\hat{o}_{b} = \hat{c}_i^{\dagger}\hat{c}_j + \hat{c}_j^{\dagger} \hat{c}_i$. The bipartite nature of the lattice implies that the two sites $i,j$ belong to two different sublattices. Since $\hat{o}_{b}^{3}=\hat{o}_{b}$ and $\hat{o}_{b}^{4}=\hat{o}_{b}^{2}=\hat{n}_{i}+\hat{n}_{j}-2\hat{n}_{i}\hat{n}_{j}$, one has 
$e^{\lambda \hat{o}_{b}}  = 1 + \sinh(\lambda)  \hat{o}_{b} + \left[\cosh (\lambda) - 1\right] \hat{o}_{b}^2$. Using these relations we can rewrite the Hamiltonian (\ref{eq:model}) into a summation of exponentials of bilinear fermion operators
%\begin{eqnarray}
%  \hat{H} &=& \sum_{b} \left(-t\hat{o}_{b} -\frac{V}{2}
%  \hat{o}_{b}^2 +\frac
%  {V}{4}\right)\nonumber  \\
%    & = &  \sum_{b} \left(\underbrace{\mystrut{2ex} \frac{-t   }{\sinh\lambda} e^{\lambda \hat{o}_{b}} }_{\hat{h}_{b}} +\frac{t}{\sinh\lambda} +  \frac{V}{4} \right)  %\nonumber\\
%%    & = & \sum_{b}\hat{h}_{b} + \mathrm{const.} 
%\end{eqnarray}
\begin{equation}
%  \hat{H} =  \sum_{b}\left(  \frac{-t }{\sinh(\lambda)}  e^{\lambda \hat{o}_{b}} + \frac{t^{2}}{V}\right), 
   \hat{H} = \frac{-t }{\sinh(\lambda)}  \sum_{b=1}^{N_{b}} e^{\lambda \hat{o}_{b}} + \mathrm{const}, 
\label{eq:exp} 
\end{equation}
where $\lambda = \ln[ {(2 t + V)}/({2 t - V}) ]$ and $N_{b}$ is the total number of bonds of the lattice. 
%Up to a constant, the Hamiltonian could be written as a sum of $\hat{h}_{b}$, which is an exponential of fermion bilinear operators acting on bonds. 
Rewriting the Hamiltonian in the form of \Eq{eq:exp} allows to trace out the exponentials of quadratic fermion terms in the Taylor series expansion of the partition function %$ Z =  \mathrm{Tr} \left(e^{- \beta \hat{H}}\right) = \sum_{n = 0}^{\infty} \frac{1}{n!} \left( \frac{\beta t}{\sinh
% \lambda} \right)^n\sum_{\{b_\ell\}}   \mathrm{Tr} \left(\prod_{\ell=1}^{n}  e^{\lambda \hat{o}_{b_{\ell}}} \right) \nonumber \\$
\begin{eqnarray}
  Z %& = & \mathrm{Tr} \left(e^{- \beta \hat{H}}\right) = \sum_{n = 0}^{\infty} \frac{(-\beta)^{n}}{n!} \Tr\left( \hat{H}^{n}\right)  \nonumber \\
%  & = & \sum_{n = 0}^{\infty} \frac{(-\beta)^n}{n!} \sum_{b_1, \ldots, b_{n}}
%  \mathrm{Tr} \left( \hat{h}_{b_{n}} \ldots \hat{h}_{b_1}\right)\nonumber \\
%   & = & \sum_{n = 0}^{\infty} \frac{1}{n!} \left( \frac{\beta t}{\sinh
%  \lambda} \right)^n\sum_{\{b_\ell\}}   \mathrm{Tr} \left(\prod_{\ell=1}^{n}  e^{\lambda \hat{o}_{b_{\ell}}} \right) \nonumber \\
  & = & \sum_{n = 0}^{\infty} \frac{1}{n!} \left( \frac{\beta t}{\sinh
  (\lambda)} \right)^n \sum_{\{b_\ell\}=1}^{N_{b}}  \det \left( I + \prod_{\ell=1}^{n}  e^{\Lambda_{b_{\ell}}} \right), 
  \label{eq:expansion}
\end{eqnarray}
where $\beta = 1/(k_{\rm B}T)$ is the inverse temperature. 
%\begin{eqnarray}
%  Z & = & \mathrm{Tr} \left(e^{- \beta \hat{H}}\right)\\
%  & = & \sum_{n = 0}^{\infty} \frac{(-\beta)^n}{n!} \sum_{\{b_\ell\}}
%  \mathrm{Tr} \left( \prod_{\ell=1}^{n} \hat{h}_{b_{\ell}}\right)\nonumber \\
%  & = & \sum_{n = 0}^{\infty} \frac{1}{n!} \left( \frac{\beta t}{\sinh
%  \lambda} \right)^n \sum_{\{b_\ell\}} \det \left( I + \prod_{\ell=1}^{n} e^{\Lambda_{b_{\ell}}} \right) \nonumber 
%\end{eqnarray}
The second summation runs over the bond indices $b_{\ell}\in[1, N_{b}]$. For each bond index $b=\braket{i,j}$, the matrix $\Lambda_{b}$ has only two nonzero elements at the two sites connected by the bond, {\it i.e.}, $(\Lambda_{b})_{pq}=\lambda(\delta_{pi}\delta_{jq}+\delta_{pj}\delta_{iq})$. 
The matrix size is $N_{s}\times N_{s}$ where $N_{s}$ is the number of lattice sites. The matrix product in \Eq{eq:expansion} is a sequence of hyperbolic rotations, in which each vertex matrix $e^{\Lambda_{b}}$ lies in the split orthogonal group~\cite{Wang:2015hm}. Physically, the matrix determinant in \Eq{eq:expansion} is the partition function of a sequence of hopping events in the imaginary time. 

The expansion (\ref{eq:expansion}) has a number of interesting properties. In the limit of $t\rightarrow0$, $\lambda\rightarrow \mathrm{i}\pi$ and ${t}/{\sinh(\lambda)}\rightarrow-{V}/{4}$. The vertex matrix is diagonal $(e^{\Lambda_{b}})_{pq} = \delta_{pq}-2\delta_{pi}\delta_{iq}-2\delta_{pj}\delta_{jq}$ and the matrix determinant in \Eq{eq:expansion} can be evaluated analytically. Since a nonvanishing matrix determinant implies that every site appears even number of times in the product, \Eq{eq:expansion} reduces to the high-temperature series expansion of the two-dimensional (2D) classical Ising model~\cite{oitmaa2006series} (as it should). More importantly, each term in the expansion is nonnegative for finite repulsive interaction strength $V/t>0$, and is thus amenable to Monte Carlo sampling. For example, when $V < 2 t$ the matrix determinant is nonnegative for any $\{b_{\ell}\}$ because the matrix product $\prod_{\ell=1}^{n}  e^{\Lambda_{b_{\ell}}}$ belongs to the identity component of the split orthogonal group~\cite{Wang:2015hm}. For $V > 2 t$ the matrix determinant flips sign between even and odd expansion orders~\cite{Wang:2015hm}.  However this sign is cancelled by $\sinh (\lambda)<0$ in the prefactor. The point $V = 2 t$ is singular in Eq.~(\ref{eq:expansion}) because $\lambda$ diverges. However, this can be solved with a slightly modified algorithm described in Appendix~\ref{appendix:splitV}. 
%\red{ How to restore the $V=0$ case ? How to sample it ? }
%An explicit calculation for a two site model is shown in the appendix. 

%%\subsection{Update}
\begin{figure}
 \includegraphics[width=\columnwidth]{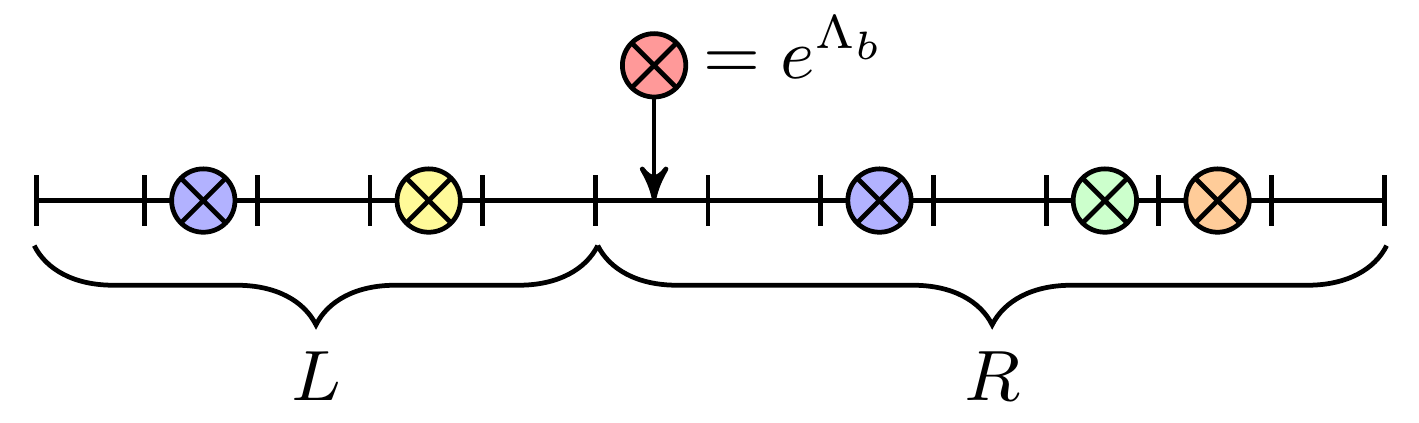}
 \caption{A Monte Carlo configuration of the truncated series expansion \Eq{eq:sse}. There are $M=12$ fixed number of slots. Each slot either holds a vertex matrix with bond type $b_{\ell}\in [1, N_{b}]$ (denoted by colored vertices) or an identity matrix if $b_{\ell}=0$ (denoted by an empty slot). 
The red vertex with an arrow indicates an insertion update. Before the update, there are $n=5$ vertices, and $L,R$ denote the partial matrix products of the embraced slots. 
}
\label{fig:sse-config}
\end{figure}

To sample the series  (\ref{eq:expansion}) we use the SSE algorithm~\cite{Sandvik:1991tn, 1999PhRvB..5914157S, Syljuasen:2002hw}. First we truncate the series expansion to a maximal expansion order $M$, and then pad identity matrices into the matrix product. This truncation is determined in the equilibration phase of the simulation, and does not introduce any bias into the simulation. Taking into account the distribution of these identity matrices in the fixed-length matrix sequence, the expansion (\ref{eq:expansion}) reads 
\begin{equation}
Z = \sum_{n = 0}^{M} \frac{(M-n)!}{M!} \left( \frac{\beta t}{\sinh
  (\lambda)} \right)^n \sum_{\{b_\ell\}=0}^{N_{b}}  \det \left( I + \prod_{\ell=1}^{M}  e^{\Lambda_{b_{\ell}}} \right),  
  \label{eq:sse}
\end{equation}
where in the second summation we extend the bond type to include identity vertex matrices $e^{\Lambda_{b=0}}=I$. Figure~\ref{fig:sse-config} shows an example of one of the configurations in this sum. Equation~(\ref{eq:sse}) has the structure of the standard SSE method, except for the appearance of the matrix determinant which motivates the use of determinantal QMC techniques for an efficient simulation. Note that, although the sequence of matrix product in \Eq{eq:sse} resembles the traditional discrete time formulation of the  determinantal QMC approach~\cite{Blankenbecler:1981vj}, there is no time-discretization error in the SSE formalism. 

To update the Monte Carlo configurations we sweep through the matrix sequence, and for each matrix either propose to change its bond index from $b_{\ell}=0$ to $b_{\ell}=b\in[1, N_{b}]$, or vice versa. Let $n$ denote the number of non-identity matrices before the update, $L=\prod_{\ell=k+1}^{M}  e^{\Lambda_{b_{\ell}}}$ and $R=\prod_{\ell=1}^{k}  e^{\Lambda_{b_{\ell}}}$ denote the partial matrix products
upto the $k$-th slot which is under consideration, shown in Fig.~\ref{fig:sse-config}. The Metropolis-Hastings~\cite{Metropolis:1953in, Hastings:1970aa} acceptance rate of the insertion update is 
%of changing a null operator $b_{\ell}=0$ to an operator with $b_{\ell}=b\in[1, N_{b}]$ with 
\begin{equation}
 p_{0\rightarrow b} = \min\left\{1,  \frac{N_b \beta t}{(M - n) \sinh (\lambda)} \frac{\det \left( I + L e^{\Lambda_{b}} R \right)}{\det (I + L R)} \right\},   
  \label{eq:add}
\end{equation}
and 
\begin{equation}
 p_{b\rightarrow 0} =  \min\left\{1,  \frac{(M - n + 1) \sinh (\lambda)}{N_b \beta t} \frac{\det \left( I +L e^{- \Lambda_{b}} R \right)}{\det (I + LR)} \right\} 
  \label{eq:removal}
\end{equation}
for the removal update. 
Conceptually, these updates are similar to the ``diagonal update'' in the standard SSE algorithm~\cite{1999PhRvB..5914157S, Syljuasen:2002hw}, but here they are actually  sufficient to ensure ergodicity. By keeping track of the Green's function $G=(I+RL)^{-1}$ and making use of the sparseness of the vertex matrix $e^{\Lambda_{b}}$, one can compute the acceptance rates Eqs.~(\ref{eq:add}) and (\ref{eq:removal}) 
in constant time. The Green's function also facilities measurements of the physical observables, similarly to conventional determinantal QMC methods~\cite{Blankenbecler:1981vj, Assaad:2008hx} and recent linear-$\beta$ scaling continuous-time quantum Monte Carlo (LCT-QMC) methods~\cite{Iazzi:2015hi, 2015PhRvB..91w5151W}. We describe implementation details of an efficient simulation in Appendix~\ref{appendix:fastupdate}. 

%Expansion of fixed length of M.
%\begin{eqnarray*}
%  Z & = & \sum_{n=0}^M \frac{(M - n) !}{M!} \left( \frac{\beta t}{\sinh
%  (\lambda)} \right)^n \sum_{b_1, \ldots b_n} \det \left( I + e^{\lambda
%  O_{b_1}} \ldots e^{\lambda O_{b_n}} \right)
%\end{eqnarray*}

Similar to standard SSE simulation~\cite{Sandvik:1991tn}, the average expansion order is related to the expectation value of the total energy
\begin{equation}
%  \braket{\hat{H}}  =  - \frac{\braket{ n }}{\beta} + N_{b} \left( \frac{t}{\sinh
%  (\lambda)} + \frac{V}{4} \right)   
  \braket{n} =  -\beta\braket{\hat{H}}  + \beta N_{b}t^{2}/V, \label{eq:expansionorder}
\end{equation}
where the last term accounts for the constant offset in \Eq{eq:exp}. The noninteracting limit is a singular point of the series expansion (\ref{eq:expansion}), i.e., the average expansion order diverges at $V=0$ even for finite systems at finite inverse temperature. Nevertheless, the present method is still advantageous in the physically interesting region $V\sim t$ where the series is well-behaved. According to \Eq{eq:expansionorder} the truncation $M$ has to grow as $\mathcal{O}(\beta N_{s})$ to accommodate these number of non-identity matrices. Combined with the $\mathcal{O}(N_{s}^{2})$ fast update of the Green's function (Appendix~\ref{appendix:fastupdate}), the present SSE algorithm exhibits an overall $\mathcal{O}(\beta N_{s}^{3})$ scaling, same as the LCT-QMC methods~\cite{Iazzi:2015hi, 2015PhRvB..91w5151W} and the traditional discrete-time algorithm~\cite{Blankenbecler:1981vj}. 

%Specifically in the SSE method, the temperature derivative of physical observables can be directly evaluated
%\begin{equation}
%\frac{d\braket{\hat{O}}}{dT} = \left\langle\frac{\partial \hat{O}}{\partial T}\right\rangle - \frac{\braket{ \hat{O} n } - \braket{\hat{O}}\braket{n}}{T}
%\end{equation}
%Specifically, for the specific heat $C_{V}= \frac{d\braket{\hat{H}}}{dT}=\braket{n^{2}}-\braket{n}^{2}-\braket{n}$. 
%Structure factor and Binder ratio

\section{Results}
\begin{figure}
 \includegraphics[width=\columnwidth]{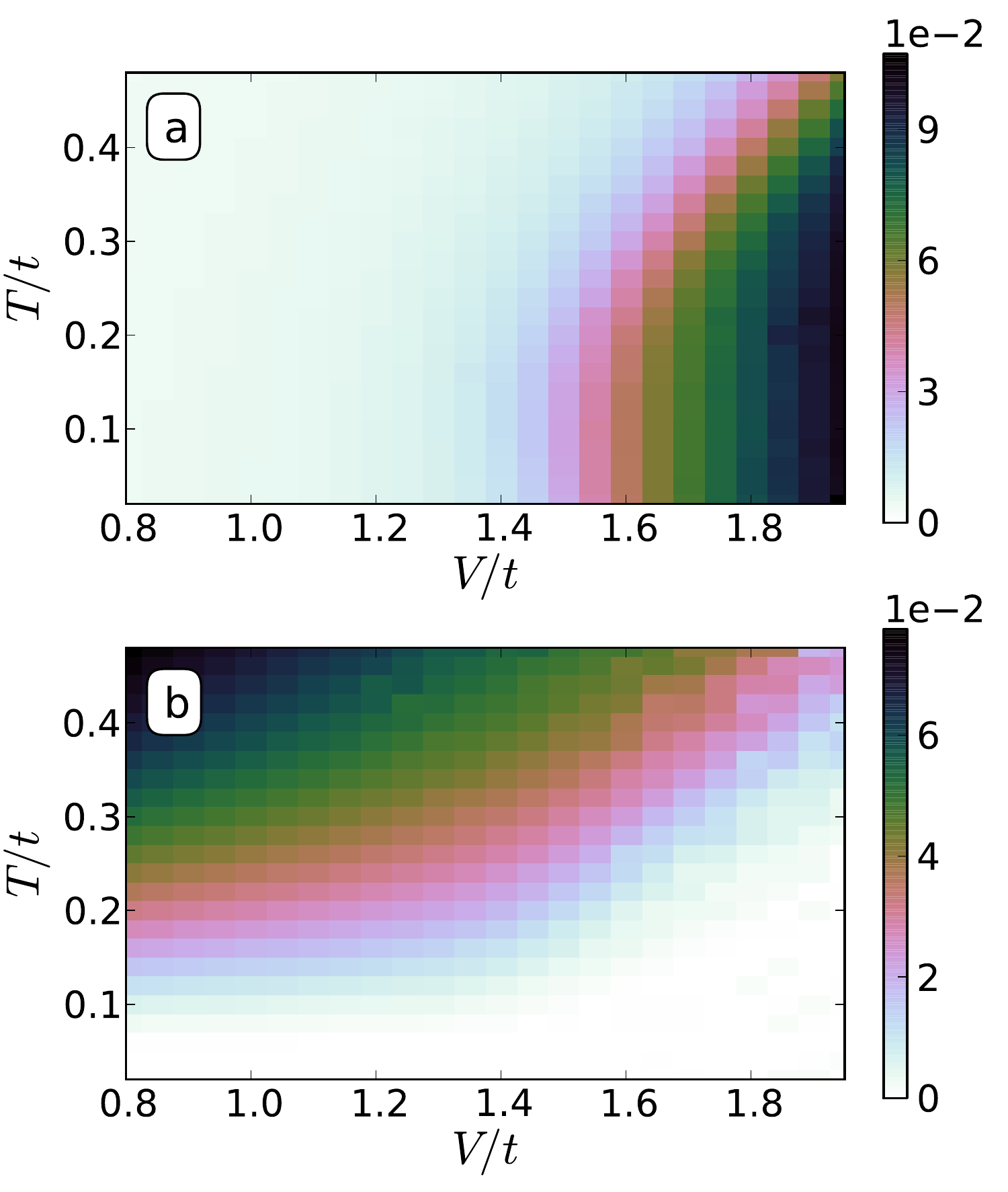}
 \caption{(a) CDW structure factor (\ref{eq:M2}) and (b) compressibility (\ref{eq:kappa}) of the model (\ref{eq:model}) on a $\mathcal{L}=10$ honeycomb lattice. }
\label{fig:scanTV}
\end{figure}
%\begin{figure*}[t]
% \includegraphics[width=\textwidth]{M2.pdf}
% \caption{Scaled CDW structure factor (\ref{eq:M2}) versus the inverse system length for various interaction strengths. At the critical temperature $M_{2}\mathcal{L}^{1/4}$ reaches a  size independent value.}
%\label{fig:M2}
%\end{figure*}

\begin{figure*}[t]
 \includegraphics[width=\textwidth]{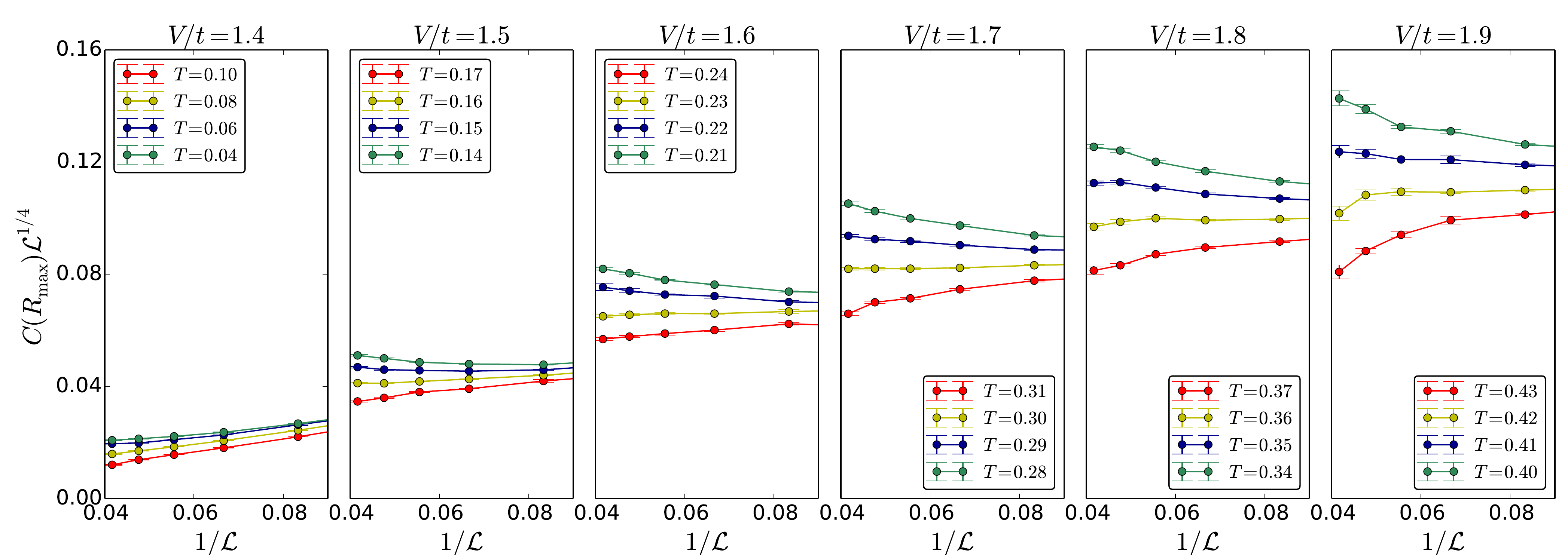}
 \caption{Scaled density correlation at the maximum distance (\ref{eq:CRmax}) versus the inverse system length, for various interaction strengths. At the critical temperature $C(R_\mathrm{max})\mathcal{L}^{1/4}$ reaches a size independent value.}
\label{fig:CRmax}
\end{figure*}

We start by discussing the general behavior of physical observables in a wide parameter range. Figure~\ref{fig:scanTV} shows the staggered CDW structure factor 
\begin{equation}
M_{2} = \frac{1}{N_{s}^{2}}  \left\langle\left(\hat{N}_{A} - \hat{N}_{B}\right)^{2} \right \rangle, \label{eq:M2}
\end{equation}
and the compressibility calculated from the total density fluctuation
\begin{equation}
%\kappa = \frac{\beta}{N}\left\langle \left[\sum_{i} \left(\hat{n}_{i}-\frac{1}{2}\right) \right]^{2} \right \rangle. \label{eq:kappa} 
%\kappa = \frac{\beta}{N}\left(\left\langle \left(\sum_{i}\hat{n}_{i} \right)^{2} \right \rangle -   \left\langle  \sum_{i}\hat{n}_{i} \right\rangle^{2} \right). \label{eq:kappa} 
\kappa = \frac{\beta}{N_{s}}\left(\left\langle \hat{N}^{2} \right \rangle -   \left\langle  \hat{N}\right\rangle^{2} \right), \label{eq:kappa} 
%C(r_\mathrm{max})&=& \frac{1}{N}\sum_{i}\left\langle\left(n_{i}-\frac{1}{2}\right)\left(n_{i+r_\mathrm{max}}-\frac{1}{2}\right)\right\rangle%\sim |i-j|^{-\eta}
\end{equation}
where $\hat{N}_{A(B)} = \sum_{i\in A(B)}\hat{n}_{i}$ is the total particle number in the $A(B)$ sublattice, and $\hat{N}=\hat{N}_{A}+\hat{N}_{B}$ is the total particle number on a honeycomb lattice with $N_{s} = 2\mathcal{L}^{2}$ sites. We have chosen $\mathcal{L}=10$ in Fig.~\ref{fig:scanTV} to avoid a finite compressibility in the weak coupling region, which is an artifact due to finite density-of-states at zero energy of the clusters with $\mathcal{L}=0\bmod 3$. The CDW structure factor $M_{2}$ shown in Fig.~\ref{fig:scanTV}(a) is related to the square of the CDW order parameter and increases in the low-temperature and strong-coupling limit (lower-right corner). In the same parameter region the compressibility shown in Fig.~\ref{fig:scanTV}(b) is suppressed because the CDW state is gapped. These features can be used to detect the CDW phase in an experimental realization. In strong coupling limit the temperature scale of the CDW transition is set by the interaction strength $V$ instead of the superexchange scale like in the Hubbard model. This makes it particularly promising in the ultracold atom setups, where the onset of the short-range magnetic correlation and suppression of compressibility have been observed for the Hubbard model~\cite{Greif:2013kb, Hart:2015exa, Duarte:2015cnb, Greif:2015wi, Cocchi:2015td}. 

We next proceed to accurately determine the critical temperature based on scaling behavior of the density correlations. At the critical temperature, they decay algebraically at large distances as
\begin{equation}
C(i, j)= \left\langle\left(\hat{n}_{i}-\frac{1}{2}\right)\left(\hat{n}_{j}-\frac{1}{2}\right)\right\rangle \sim \frac{1}{|i-j|^{\eta}} \label{eq:CRmax}, 
\end{equation}
where $|i-j|$ denotes the distance between the sites $i,j$ and $\eta=1/4$ is the critical exponent of 2D Ising transition. Away from the critical temperature, the correlation function decays exponentially to either zero in the disordered phase or a finite value in the ordered phase (which is the square of the CDW order parameter). 
%Therefore, the quantity $C(R_\mathrm{max})\mathcal{L}^{\eta}$ approaches to a size independent value in large system limit. scales the density correlation between two sites at maximum distance 
%\begin{equation}
%C(r)=  \frac{1}{N_{s}}\sum_{i}\left\langle\left(n_{i}-\frac{1}{2}\right)\left(n_{i+r}-\frac{1}{2}\right)\right\rangle \sim |i-j|^{-\eta}
%\end{equation}
%$C(|i-j|)= \braket{\left(n_{i}-\frac{1}{2}\right)\left(n_{j}-\frac{1}{2}\right)}\sim |i-j|^{-\eta}$ 
Figure~\ref{fig:CRmax} shows the scaled density correlation function at the largest distance $R_\mathrm{max}=4\mathcal{L}/3$ on the honeycomb lattice
%In Fig.~\ref{fig:M2} we show the scaled CDW structure factor versus inverse system size 
for $\mathcal{L}=12,15,18,21,24$ and various interaction strengths. At the critical temperature, the scaled density correlation $C(R_\mathrm{max})\mathcal{L}^{\eta}$ reaches a size independent value as $\mathcal{L}$ grows. %Figure \ref{fig:Binder} shows the Binder ratios calculated for various interaction strength. 
Based on this we determine the critical temperatures for various interaction strengths shown in phase diagram of Fig.~\ref{fig:Phasediag}. The error bars on the critical temperature indicate the upturn and downturn of the scaled density correlations in Fig.~\ref{fig:CRmax}.%structure factor $M_{2}\mathcal{L}^{\eta}$.

%Figure~\ref{fig:Phasediag} shows the critical temperature of the model (\ref{eq:model}) on a honeycomb lattice. %The red dot marks the previously determined location of the quantum critical point~\cite{Wang:2014iba, 2015PhRvB..91w5151W}. 
We see that in the strong-coupling limit $V\gg t$, the critical temperature asymptotically approaches the dashed black line $T_{c}=0.38V$, corresponding to the classical Ising model on the honeycomb lattice~\cite{1967RPPh...30..615F}. Quantum fluctuations substantially suppress the critical temperature away from this strong coupling limit and the critical temperature drops to zero at the quantum critical point between the CDW and the Dirac semimetal state~\cite{Wang:2014iba,2015PhRvB..91w5151W,Li:2015cwb}. Fitting the critical temperature around the quantum critical region to the form $T_{c}=A(V-V_{c})^{\nu z}$ gives $V_{c}/t = 1.36(3), \nu z = 0.72(9)$. Extracting quantum critical properties in this way is certainly indirect, but it provides a consistency check against previous results~\cite{Wang:2014iba, 2015PhRvB..91w5151W, Li:2015cwb}. Setting $z=1$ due to the relativistic invariance of the model, these estimates are consistent with the ground-state LCT-QMC results $V_{c}/t=1.356(1),\nu=0.80(3)$~\cite{2015PhRvB..91w5151W, Wang:2014iba} and also the results of~\cite{Li:2015cwb}. Along the transition line, the nature of the phase transition undergoes a crossover from a fermionic quantum critical point to a 2D Ising phase transition. 

\section{Discussion and Outlook}
%We presented a way to apply the stochastic series expansion algorithm to fermionic models in higher dimensions. The key step is to rewrite the Hamiltonian as a sum of exponentials of fermion bilinear operators (such as in \Eq{eq:exp}) and trace out fermions in the Taylor series expansion of the partition function. The techniques of determinantal QMC can then be used to eliminate the fermion sign problem and perform efficient simulation. 
The proposed SSE approach applies as well to the ground-state projection scheme~\cite{Sandvik:2005je}. The approach also allows for easy computation of quantum information quantities such as the fidelity susceptibility~\cite{Wang:2015bva} and the R\'enyi entanglement entropy~\cite{Grover:2013cs, Broecker:2014fl}. 

%Using this method we mapped out the finite-temperature phase diagram of the spinless $t$-$V$ model on the honeycomb lattice as shown in Fig.~\ref{fig:Phasediag}. 
In principle, the presented calculations for the $t$-$V$ model could have been performed using the $\mathcal{O}(\beta N_{s}^{3})$ scaling LCT-QMC methods~\cite{Iazzi:2015hi, 2015PhRvB..91w5151W} or the traditional discrete-time method~\cite{Li:2015jf, Li:2015cwb}, or even the continuous-time interaction expansion method with a suboptimal $\mathcal{O}(\beta^{3}N_{s}^{3})$ scaling~\cite{Huffman:2014fj, Wang:2014iba}. Our SSE approach shows the best performance of all available methods in the vicinity of the quantum critical point. The SSE implementation requires fewer numerical operations, despite its larger expansion order compared to the continuous-time interaction expansion methods~\cite{1999PhRvL..82.4155R, Rubtsov:2005iw, Iazzi:2015hi, 2015PhRvB..91w5151W}, see Appendix~\ref{appendix:splitV} and~\ref{appendix:fastupdate} for detailed discussions. Simulations presented in this paper have reached a maximum cutoff $M\approx150,000$ for the $\mathcal{L}=24$ lattice at $V/t=1.4$ and $\beta t=25$. As a comparison, to match this performance in the traditional discrete-time approach~\cite{Blankenbecler:1981vj}, one needs to use a (too large) time step $\Delta \tau t=0.3$, to have a comparable number of auxiliary fields $N_{b}\beta/\Delta{\tau}=144,000$ to sum up. 

%application range
Our work points to several interesting possibilities. One may wonder whether is it possible to apply this hybrid SSE/determinantal approach to a broader range of fermionic models. It may seem that the rewriting in \Eq{eq:exp} puts a rather strong constraint on the type of Hamiltonians. However, using the approach of Refs.~\cite{Rombouts:1998tf, 1999CoPhC.121..446R} it is possible to decompose the most general form of two-body interactions (including quantum-chemistry  Hamiltonians) into a summation of exponentials of fermion bilinear terms. The difficulty is  avoiding the fermion sign problem. For the specific case of the spinless $t$-$V$ model the sign problem is completely eliminated by using the split orthogonal group property of the fermionic determinant~\cite{Wang:2015hm}. For other problems that are known to be sign-problem free (such as half-filled repulsive Hubbard model on bipartite lattices) in the determinantal QMC methods, it is as yet unclear how to devise a similar SSE approach. Unlike  determinantal QMC methods~\cite{Blankenbecler:1981vj, 1999PhRvL..82.4155R, Rubtsov:2005iw, Iazzi:2015hi, 2015PhRvB..91w5151W}, here the key is a proper treatment of the single-particle hopping terms. Nevertheless, even if a sign-free simulation is not possible in general, the sign problem in the hybrid determinantal/SSE approach may still be less severe than the direct application of the standard SSE algorithm. Since by tracing out the fermions many of the fermion exchange processes are taken into account by the matrix determinant. 

Specific to the spinless $t$-$V$ model considered in this paper, it will be interesting to see whether there exists an even more efficient $\mathcal{O}(N_{s})$ scaling algorithm by utilizing the special properties of the matrix determinant in \Eq{eq:expansion}. If possible, the resulting method will be as efficient as the meron-cluster approach~\cite{Anonymous:xaWVK-gC} which only applies to $V\ge 2t$.  Finally, solving fermionic problem in the SSE framework also makes one wonder whether it is possible to construct nonlocal Monte Carlo updates for fermionic Hamiltonians~\cite{1999PhRvB..5914157S, Syljuasen:2002hw, Alet:2005cv}. 

%Finally, the rewriting \Eq{eq:exp} used in this paper and more generally in \cite{Rombouts:1998tf, 1999CoPhC.121..446R} are likely to be useful for the digital simulation of Hamiltonian dynamics on quantum computers~\cite{PhysRevLett.114.090502}. 

We end by noting an  independent study of the model (\ref{eq:model}) by Hesselmann and Wessel~\cite{HesselmannWessel} using the continuous-time interaction expansion method~\cite{Rubtsov:2005iw, Huffman:2014fj, Wang:2014iba}. %Besides determined critical temperature of the thermal phase transitions, they also carefully examined critical exponents of the fermionic quantum critical point. 
When there is an overlap, our results are in full agreement with theirs.

\section{Acknowledgment}
We thank Fakher Assaad and Stefan Wessel for helpful discussions. Simulations were performed on the M\"onch cluster and Brutus cluster of ETH Zurich. We have used ALPS libraries~\cite{BBauer:2011tz} for Monte Carlo simulations and data analysis. This work was supported by ERC Advanced Grant SIMCOFE and by the Swiss National Science Foundation through the National Center of Competence in Research Quantum Science and Technology QSIT.

\bibliographystyle{apsrev4-1}
%\bibliography{/Users/wanglei/Documents/Papers2/papers}
\bibliography{det-SSE}

%\begin{thebibliography}{99}
%\bibitem{} Kane et al. Quantum Spin Hall Effect in Graphene. Phys. Rev. Lett. (2005) vol. 95 (22) pp. 4
%\end{thebibliography}

\appendix

\section{An improved algorithm for $V\sim2t$\label{appendix:splitV}}
The algorithm presented above does not apply to the case  $V=2t$ because $\lambda=\ln\left( \frac{2 t + V}{2 t - V} \right)$ diverges at this point. Moreover, the vertex matrices will have large condition number and impair the numerical stability of the simulation even at $V\sim2t$. Here we present an improved algorithm to solve these two problems. In practice, only minor modifications to the original algorithm are needed.

The solution is to split  the interaction term into two parts $V = V_1 + V_2$ and treat them separately. The first part is combined with the hopping term as was done in \Eq{eq:exp}, while the second part is written as an exponential of fermion bilinear by its own~\cite{2015PhRvB..91w5151W, Wang:2015hm}, 

\begin{equation}
  \hat{H} %& = & \sum_{b} - t (c_i^{\dagger} c_j + c_j^{\dagger} c_i) + V_1 
%  \left( n_i n_j - \frac{n_i + n_j}{2} \right) + V_2  \left( n_i n_j -
%  \frac{n_i + n_j}{2} \right)\\
=  \sum_{b=1}^{N_{b}} \left( \frac{-t }{\sinh(\lambda_{1})} e^{\lambda_{1} \hat{o}_{b}}+ \frac{V_{2}}{4}e^{\lambda_{2} \hat{o}_{b}} \right)+\mathrm{const} \label{eq:V1V2},  
%  & = & \sum_{< i, j >} - t \left[ \frac{1}{\sinh (\lambda)} + \hat{O}_{i j}
%  + \frac{V}{2 t} \hat{O}_{i j}^2 \right] + \sum_{< i, j >}  \left[
%  V_2 \left( n_i n_j - \frac{n_i + n_j}{2} \right) - \Gamma \right] + \sum_{<
%  i, j >} \left( \frac{t}{\sinh (\lambda)} + \Gamma \right)\\
%  & = & \sum_{< i, j >} \frac{- t}{\sinh (\lambda)} e^{\lambda \hat{O}_{}} +
%  \sum_{< i, j >}  \frac{- \Gamma}{2} \sum_{\sigma = \pm} e^{\sigma \gamma
%  \hat{O}_{i j}} + \sum_{< i, j >} \left( \frac{t}{\sinh (\lambda)} + \Gamma
%  \right)\\
%  & = & \sum_{< i, j >} \frac{- t}{\sinh (\lambda)} e^{\lambda \hat{O}_{}} +
%  \sum_{< i, j >}  \frac{V_2}{4} e^{i \pi \hat{O}_{i j}} + \sum_{< i, j >}
%  \left( \frac{t}{\sinh (\lambda)} - \frac{V_2}{4} \right)\\
%  & = & - \sum_{< i, j >} \sum_{\alpha} f_{\alpha} e^{\lambda_{\alpha}
%  \hat{O}} + \sum_{< i, j >} \left( \frac{t}{\sinh (\lambda)} - \frac{V_2}{4}
%  \right)\\
%  &  & \\
%  f_0 & = & \frac{t}{\sinh (\lambda)} = \frac{t^2 - 0.25 V_1^2}{V_1}\\
%  f_1 & = & - \frac{V_2}{4}\\
%  \lambda_0 & = & \log \left( \frac{2 t + V_1}{2 t - V_1} \right)\\
%  \lambda_1 & = & i \pi\\
%  &  & \\
%  \Gamma - V_2 \left( n_i n_j - \frac{n_i + n_j}{2} \right) & = &
%  \frac{\Gamma}{2} \sum_{\sigma} e^{\sigma \gamma \hat{O}}\\
%  \gamma & = & \mathrm{acosh} \left( 1 + \frac{V_2}{2 \Gamma} \right)
\end{equation}
where $\lambda_{1} = \ln\left( \frac{2 t + V_{1}}{2 t - V_{1}} \right)$ and $\lambda_{2}=\mathrm{i}\pi$. Such a splitting doubles the total number of bond types. Including the identity matrices, the truncated series expansion reads
\begin{eqnarray}
  Z %& = &\mathrm{Tr} \left[ e^{\beta \sum_{\alpha} f_{\alpha} \exp
 % (\lambda_{\alpha} \hat{O})} \right]\\
  =  \sum_{n = 0}^{M} \frac{(M-n)!}{M!} \left(\frac{\beta t}{\sinh(\lambda_{1})}\right)^{n_{1}} \left(\frac{-\beta V_{2}}{4}\right)^{n_{2}} \nonumber \\ \times \sum_{\{b_{\ell}\}=0}^{2N_{b}} \det \left( I + \prod_{\ell=1}^{M}e^{\Lambda_{b_{\ell}}} \right)
  \label{eq:splitV}
\end{eqnarray}
where $n=n_{1}+n_{2}$ is the total number of non-identity matrices. There are $n_{1}$ bonds with $b_{\ell}\in [1,N_{b}]$ and $n_{2}$ bonds with $b_{\ell}\in [N_{b}+1, 2N_{b}]$. In the matrix ${\Lambda_{b}}$  nonzero matrix elements may now be either $\lambda_{1}$ or $\lambda_{2}$ depending on the bond type. The acceptance rates Eq.~(\ref{eq:add}) and (\ref{eq:removal}) 
need to be modified accordingly. Given $V_{1}>0,V_{2}>0$ the splitting does not introduce sign problem since each term of \Eq{eq:splitV} is nonnegative thanks to the split orthogonal group condition~\cite{Wang:2015hm}. 

Splitting of the terms in \Eq{eq:V1V2} provides a useful tuning knob to the algorithm. By choosing $V_{1}\neq2t$ one can avoid the singularity at $V=2t$. Moreover, the vertex matrices are better conditioned at smaller $V_{1}/t$ because $\lambda_{1}$ is smaller. The simulation is thus more stable. On the other hand, since 
\begin{equation}
%\braket{H}= -\frac{\braket{n}}{\beta}+ N_{b}\left(\frac{t}{\sinh(\lambda_{1})}+\frac{V_{1}}{4}\right)
  \braket{n} =  -\beta\braket{\hat{H}}  + \beta N_{b}t^{2}/V_{1}, 
\end{equation}
the price to pay is a larger average expansion order at smaller $V_{1}$. For all results presented in this paper, we use $V_{1}=0.5t$ for a balanced performance. 

The averages  $\braket{n_{1}}$ and $\braket{n_{2}}$ are related to the relative weight of the two terms in the Hamiltonian \Eq{eq:V1V2}. To optimize performance, we adjust the propose probability of the bond type in insertion updates according to the measured $\braket{n_{1}}$ and $\braket{n_{2}}$ in the equilibration phase. 

\section{Fast-update formula\label{appendix:fastupdate}}
For completeness, we include the formula for efficient manipulation of the matrices. These are standard techniques ported from the determinantal QMC methods~\cite{Blankenbecler:1981vj, Assaad:2008hx}.  The following techniques were used in recent LCT-QMC simulations~\cite{2015PhRvB..91w5151W, Liu:2015kx}. 

The determinant ratio in the acceptance rate \Eq{eq:add} can be calculated in terms of the Green's function $G=(I+RL)^{-1}$,  
\begin{eqnarray}
 \frac{\det \left( I + L e^{\Lambda_{b}} R \right)}{\det (I + L R)} & = & \det\left[I + ( e^{\Lambda_{b}}-I)(I-G)\right]   \label{eq:detratio} \\ 
% &=& \det\left[I + \mathcal{PP}^{T}(e^{\pm\Lambda_{b}}-I)\mathcal{PP}^{T} (I-G)\right]  \nonumber \\
&=&\det\left[I + \mathcal{P}^{T}(e^{\Lambda_{b}}-I)\mathcal{PP}^{T} (I-G)\mathcal{P} \right],  
  \nonumber 
\end{eqnarray}
where $\mathcal{P}$ is a $N_{s}\times 2$ matrix that projects to the two sites connected by the bond $b=\braket{i,j}$. In the second line we used the identity $(e^{\Lambda_{b}}-I)=\mathcal{PP}^{T}(e^{\Lambda_{b}}-I)\mathcal{PP}^{T}$ and moved the first $\mathcal{P}$ to the end using the cyclic property in such matrix determinant. Finally the determinant ratio is evaluated as a $2\times2$ matrix determinant. 

If the move is accepted, we update the Green's function using the Woodbury matrix identity,   
\begin{eqnarray}
  G^{\prime} & = & \left( I + e^{\Lambda_{b}}R L \right)^{- 1}  \label{eq:updateG} \\
  & = & { G} - {G\mathcal{P}}
 \left\{\frac{1}{\mathcal{P}^T \left[( e^{\Lambda_{b}} - I)^{- 1}
  +(I - G) \right]\mathcal{P}} \right\} 
   \mathcal{P}^{T} (I-G). \nonumber 
\end{eqnarray}
The matrix in the curly braces is of size $2\times2$. It is multiplied from the left by a $N_{s}\times 2$ matrix, and from the right by a $2\times N_{s}$ matrices. So overall the update of $G$ can be done with $\mathcal{O}(N_{s}^{2})$ operations. The removal update is implemented similarly by replacing the vertex matrix with its inverse $e^{-\Lambda_{b}}$ in Eqs.~(\ref{eq:detratio}, \ref{eq:updateG}). 

Furthermore, when we sweep through the matrix sequence, $G$ is updated by a similarity transformation like the standard determinantal QMC method~\cite{Blankenbecler:1981vj, Assaad:2008hx}. Using the sparseness of the vertex matrix, this can be done in $\mathcal{O}(N_{s})$ operations. Since there is no need to rotate to the eigenbasis of the single-particle Hamiltonian, both the calculation of the determinant ratio \Eq{eq:detratio} and the update of $G$ (\Eq{eq:updateG}) are more efficient than the corresponding calculations in the LCT-QMC methods~\cite{Iazzi:2015hi, 2015PhRvB..91w5151W}. 

In principle, keeping track of $G$ instead of the partial matrix products $L$ and $R$ is sufficient for the simulation. In the practical implementation, however, we still store the results of the singular value decomposition of $L$ and $R$, and use them to recompute the Green's function periodically~\cite{2015PhRvB..91w5151W}. These stabilization steps cost $\mathcal{O}(\beta N_{s}^{3})$ operations. 
%For the stabilization purpose we sweep through the matrix sequence back-and-force but not in a cyclic fashion like in the standard SSE implementation.  
\end{document}